\documentclass[twocolumn,showpacs,preprintnumbers,amsmath,amssymb,prl]{revtex4}

\usepackage{graphicx}
\usepackage{dcolumn}
\usepackage{bm}

\begin{document}

\title{Highly Anisotropic Superconducting Gap in Nematically Ordered and Tetragonal Phases of FeSe$_{1-x}$S$_x$ } 

\author{Y.\,Sato$^{1}$}
\author{S.\,Kasahara$^1$}
\author{T.\,Taniguchi$^{1}$}
\author{X.\,Z.\,Xing$^{1}$}
\author{Y.\,Kasahara$^1$}
\author{Y.\,Tokiwa$^{1,2}$}
\author{T.\,Shibauchi$^3$}
\author{Y.\,Matsuda$^1$}

\affiliation{$^{1}$ Department of Physics, Kyoto University, Kyoto 606-8502 Japan}
\affiliation{$^2$ Center for Electronic Correlations and Magnetism, Institute of Physics, Augsburg University, 86159 Augsburg, Germany} 
\affiliation{$^3$ Department of Advanced materials Science, University of Tokyo, Chiba 277-8561, Japan} 
\date{\today}%

\begin{abstract}
FeSe has a unique ground state in which superconductivity coexists with a nematic order without long-range magnetic ordering at ambient pressure.  Here, to study how the pairing interaction evolves with nematicity, we measured the thermal conductivity and specific heat of FeSe$_{1-x}$S$_x$, where the nematicity is suppressed by isoelectronic sulfur substitution.  We find that in the whole nematic ($0\leq x \leq 0.17$) and tetragonal ($x=0.20$) regimes, the application of small magnetic field causes a steep increase of both quantities.  This indicates the existence of deep minima or line nodes in the superconducting gap function, implying that the pairing interaction is significantly anisotropic in both the nematic and the tetragonal regimes.  Moreover, the present results indicate that the position of gap minima/nodes in the tetragonal regime appears to be essentially different from that in the nematic regime.  These results place  an important constraint on current theories.

\end{abstract}



\maketitle

Spin fluctuations are widely discussed as a primary driving force of various unconventional superconductors, whereas in iron-based superconductors, spin and orbital degrees of freedoms are closely intertwined because of the multiple $d$-orbital characters at the Fermi level~\cite{Chubukov16, Hirschfeld11}. In most iron-based superconductors, tetragonal-orthorhombic structural (nematic) and magnetic transition lines follow closely each other. These orders have been suggested to play crucial roles in superconductivity and thus strong spin and/or orbital fluctuations have been proposed to mediate the pairing~\cite{Mazin08,Kuroki08,Kontani10}. However, despite tremendous efforts in the past years, elucidating the exact pairing mechanism still remains a great challenge. 

The iron-chalcogenide superconductor FeSe~\cite{Hsu08}, comprised only of Fe-Se layer, offers a novel platform to investigate the pairing mechanism of iron-based superconductors, because it displays several remarkable properties. The superconducting transition temperature of $T_c\sim 9$\,K  dramatically increases up to 38\,K by the application of hydrostatic pressure~\cite{Sun16}. The superconductivity at ambient pressure coexists with a nematic order, whose properties are distinctly different from the other iron-based superconductors. The nematic transition occurs at $T_s\approx 90$\,K,  which is accompanied by the splitting of the $d_{xz}$ and $d_{yz}$ orbits ($\Delta E=E_{yz}-E_{xz}\approx 60$\,meV)~\cite{Nakayama14,Shimojima14,Watson15,Suzuki15}. Although $T_s$ is comparable to other iron-based superconductors, no sizable low-energy spin fluctuations are observed above $T_s$ and no long-range magnetic order occurs below $T_s$ at ambient pressure~\cite{McQueen09, Imai09, Boehmer15, Baek15}. These results have put questions into the spin fluctuation scenario envisaged in other iron-based superconductors.  Although there is argument that the magnetic fluctuation mechanism is still applicable~\cite{Chubukov15,Wang15,Yu15}, alternative scenario where fluctuations stemming from orbital degree of freedom play a primary role has aroused a great interest~\cite{Boehmer15,Baek15,Massat16,Yamakawa16PRX,Onari16}. 


As shown in Fig.\,1(a), the Fermi surface in the nematic phase consists of an elliptical hole pocket at the Brillouin zone center (h1), elongated along $\Gamma$ - $M_y$ line, and  compensated electron pockets near the zone boundary (e1 and e2)~\cite{Onari16}.  
It has been reported that e1 pocket is divided into two Dirac-like electrons in the presence of large orbital splitting~\cite{Tan16,Zhang16,Yamakawa16}, although the detailed structure of the Fermi surface is still controversial~\cite{Watson16,Watson17}.  
The size of all the pockets is extremely small, occupying only 1\,-\,3\% of the whole Brillouin zone~\cite{Kasahara14, Terashima14, Watson15b}. 
Since the superconducting gap structure is intimately related to the pairing interaction, its elucidation is crucially important. The superconducting gap of FeSe has been reported to be highly anisotropic with deep minima or line nodes~\cite{Kasahara14, Song11, Hope16}. Recent angle resolved photoemission spectroscopy (ARPES) and the quasiparticle interference (QPI) obtained from scanning tunneling microscopy (STM) measurements consistently report the gap minima or nodes located near the vertices  along the major axis ($\Gamma$ - $M_y$ direction)  of elliptical hole pockets~\cite{Xu16,Hashimoto17,Sprau16}, although the gap structure of the electron pockets is less clear.   

The large anisotropy of the superconducting gap in FeSe is highly unusual because it directly implies that the pairing interaction strongly depends on the position of a tiny Fermi surface. However, the relationship between the nematicity and pairing interaction remains largely elusive.  To tackle this key issue, it is of primary importance to clarify how the nematicity affects on the superconducting gap structure.  Recently, it has been reported that isoelectronic sulfur substitution in FeSe strongly suppresses the nematic transition~\cite{Abdel-Hafiez15, Moore15, Watson15c}, leading to  the appearance of tetragonal regime~\cite{Hosoi16}, whose band structure is shown in Fig.\,1(b)~\cite{Onari16}. The $T$\,-\,$x$ phase diagram of FeSe$_{1-x}$S$_x$ is depicted in Fig.\,1(c).   At $x_c \approx$ 0.17, $T_s$ is suppressed to zero [nematic quantum critical point (QCP)]~\cite{Hosoi16}. As $x$ is increased in the nematic regime, $\Delta E$ is suppressed and elliptical hole pocket becomes more circular while keeping its volume nearly constant~\cite{Watson15c}.  FeSe$_{1-x}$S$_x$, therefore, offers a fascinating opportunity to investigate the nature of the pairing interaction. 

In this Letter, we report the anisotropy of the superconducting gap in FeSe$_{1-x}$S$_x$ single crystals in a wide $x$ range from the nematic to tetragonal regime beyond the nematic QCP. We measured the thermal conductivity $\kappa$ down to 90\,mK and specific heat $C$ to 400\,mK in magnetic field up to $\mu_0H=14$ T. We find that both quantities reveal that the gap is highly anisotropic, which has deep minima or line nodes, in the whole nematic regime for $0\leq x \leq x_c$, indicating that the gap anisotropy is not directly related with the nematicity. Moreover, the gap is also highly anisotropic in the tetragonal regime. These imply that the pairing interaction is highly anisotropic in both the nematic and the tetragonal regimes.  Furthermore, the position of gap minima/nodes in the tetragonal regime appears to be essentially different from that in the nematic regime.  We discuss these results in the light of orbital dependent pairing scenario. 

\begin{figure}[t]
  \begin{center}
    \includegraphics[width=1.0\linewidth]{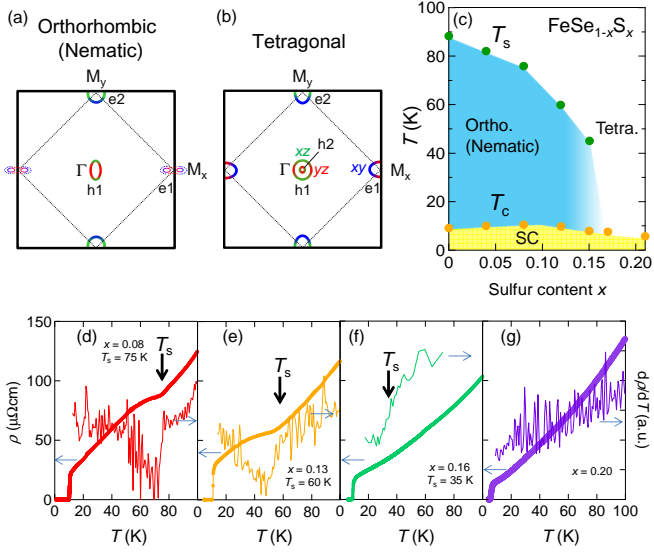}
  \end{center}
  \caption{ (a), (b) Schematic illustrations of the Fermi surface in the nematic and the tetragonal phases \cite{Onari16}. Green, red, and blue areas represent the Fermi surface regimes dominated by $d_{xz}$, $d_{yz}$ and $d_{xy}$ orbital characters, respectively. (c) $T$-$x$ phase diagram of FeSe$_{1-x}$S$_x$. (d)-(g) Temperature dependence of resistivity and its temperature derivative for single crystals of FeSe$_{1-x}$S$_x$ for (d) $x = 0.08$, (e) 0.13, (f) 0.16, and (g) 0.20.}
  \label{fig:Fig1.eps}
\end{figure}

Single crystals of FeSe$_{1-x}$S$_x$ ($x=0$, 0.08, 0.13, 0.16 and 0.20) were grown by chemical vapor transport technique~\cite{Boehmer13, Hosoi16}.  
Observation of quantum oscillations, even in the heavily substituted sample with $x=0.2$~\cite{Kasahara17, Coldea16}, nearly 100\,\% Meissner signal, and sharp jump in specific heat all demonstrate high quality of the samples. 
Specific heat was measured for $x=0$, 0.08, 0.13 and 0.20 by the quasi-adiabatic method in $^3$He cryostat.    The thermal conductivity was measured on the crystals with the same $x$ values by the standard steady-state method by applying the thermal current in the two-dimensional (2D)-plane in a dilution refrigerator. In addition to these crystals, we measured $\kappa$ for $x=0.16$ in the vicinity of nematic QCP. Since the physical properties of the crystals near QCP are sensitive to the inhomogeneous distribution of sulfur, we carefully selected a tiny crystal with a sharp superconducting transition.  For both $C$ and $\kappa$ measurements, we applied magnetic field perpendicular to the 2D plane ({\boldmath $H$} $\parallel c$). 
 
Figures\,1(d), (e), (f) and (g) depict the $T$-dependences of the resistivity $\rho$ and  $d\rho/dT$ for $x=0.08$, 0.13, 0.16 and 0.20, respectively. The nematic transition temperatures determined by the jump of $d\rho/dT$ are $T_s \approx 75$, 60 and 35\,K for $x=0.08$, 0.13 and 0.16, respectively. These values are consistent with the previous report~\cite{Hosoi16}.  At $x = 0.20$, no anomaly is observed in $d\rho/dT$, indicating that the system is in the tetragonal regime. $T_c$ and the superconducting upper critical field $H_{c2}$ for $x=0.20$ are reduced from those of the crystals in the nematic regime.

 \begin{figure}[t]
 	\begin{center}
 		\includegraphics[width=1.0\linewidth]{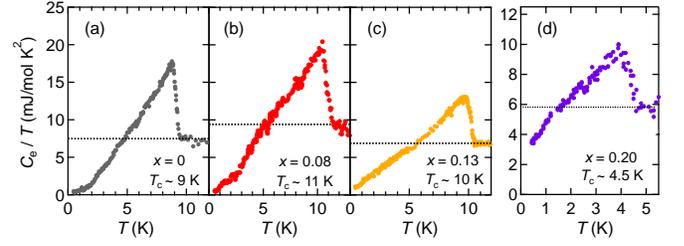}
 	\end{center}
 	\caption{The electronic component of the specific heat divided by temperature, $C_e/T$ vs. $T$ in FeSe$_{1-x}$S$_x$ for (a) $x = 0$, (b) 0.08, (c) 0.13, and (d) 0.20, respectively. }
 	\label{fig:Fig2.eps}
 \end{figure}


Figures\,2(a), (b), (c) and (d) show the $T$-dependences of the electronic component of specific heat divided by temperature, $C_e/T$ for $x=0$, 0.08, 0.13 and 0.20, respectively.  
We obtained $C_e$ by subtracting the change of $C_n$($\mu_0H=14$\,T) in the normal state from the value at $T_c$, $C_{e}=C(T)-\Delta C_n$, $\Delta C_n = C_n(T)-\gamma T_c$. 
For $x=0$, 0.08, 0.13 and 0.20, the values of $\mu_0H_{c2}$ are approximately 16, 20, 18, and 3\,T, respectively. 
Since $\mu_0H_{c2}(T)$ in the nematic regime exceeds the maximum field of our experimental setup, 14\,T, at low temperatures, $C_n(T)$ below $T_c(14\,{\rm T})$ is estimated by extrapolating a curve obtained by the fitting of $C$ above $T_c$ with $C_n(T)=\gamma T+\beta T^3+A_5T^5$. 
At $T_c$, $C_e/T$ exhibits a sharp jump for all $x$, showing good homogeneity of S substitution. The Sommerfeld coefficient $\gamma$ is  7 - 9 \,mJ/mol\,K$^2$ for all crystals in the nematic regime, suggesting that the electron correlation is little influenced by S substitution.   
The ratio of specific heat jump and normal state specific heat, $\Delta C_e/\gamma T_c$ = 1.5 for $x=0$, is larger than the BCS value of 1.43, while for 0.08, 0.13 and 0.2,  $\Delta C_e/\gamma T_c = 1.3$, 1.1, and 0.73, respectively, are smaller than the BCS value.  
This may be due to the multigap nature of the superconductivity.  For $x=0.20$, $C_e/T$ below $T_c$ shows a concave-downward curvature, which also supports the multigap superconductivity.

Figures\,3(a), (b), (c) and (d) show the $H$-dependences of $C/T$ at around 450\,mK for $x=0$, 0.08, 0.13 and 0.20, respectively.  
In conventional fully gapped superconductors, $C(H)/T$ increases linearly with $H$ due to the induced quasiparticles inside vortex cores. In stark contrast, as shown in the insets of Figs.\,3(a)-(d), $C(H)/T$ increases with $\sqrt{H}$ for all $x$ at low fields. In superconductors with highly anisotropic gap, the Doppler shift of the delocalized quasiparticle spectrum induces remarkable field dependence of density of states with $\sqrt{H}$-dependence for line node. 
For $x=0$, 0.08, and 0.13 in the nematic regime, $C(H)/T$ deviates from the $\sqrt{H}$-dependence at $H^*$ shown by arrows. For $x=0.08$ and 0.13, $C(H)/T$ exhibits a kink at $H^*$.  Above $H^*$, $C(H)/T$ increases slowly as $C(H)/T\propto H^\alpha$ with $\alpha\agt$1. The slight upward curvature of $C(H)/T$ above $H^*$  for $x=0$ and 0.13 is attributed to the Pauli paramagnetic effect on the superconductivity \cite{Tsutsumi15}. The initial steep increase of $C(H)/T$ below $H^*$ indicates that substantial portion of the quasiparticles is already restored at magnetic field much below $H_{c2}$. The slope change at $H^*$ provides evidence for multigap superconductivity; $H^*$ is interpreted as a virtual upper critical field that determines the $H$-dependence of the smaller gap. The $\sqrt{H}$-behavior below $H^*$ indicates the presence of Fermi pocket, whose superconducting gap is small and highly anisotropic with line node or deep minima.  Moreover, $H^\alpha$-dependences with $\alpha\agt1$ above $H^*$ indicate the presence of another Fermi pocket, whose gap is much larger and isotropic.

 \begin{figure}[b]
 	\begin{center}
 		\includegraphics[width=0.98\linewidth]{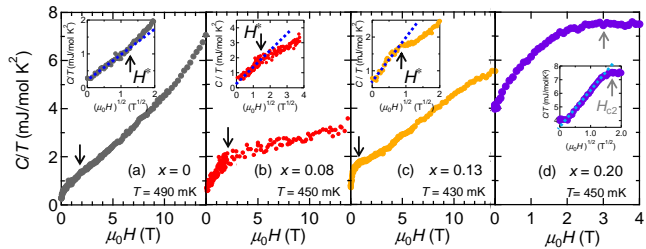}
 	\end{center}
 	\caption{Magnetic field dependence of $C/T$ for (a) $x = 0$, (b) 0.08, (c) 0.13, and (d) 0.20.  Insets show the same data plotted as a function of $\sqrt{\mu_0 H}$. $H^*$ represents a magnetic field at which $C/T$ deviates from $\sqrt{H}$-dependence. The black arrows in the main panel indicate $H^*$.  The gray arrow in (d) indicates upper critical field.}
 	\label{fig:Fig3.eps}
 \end{figure}

  For $x=0.20$ in the tetragonal regime,  $\sqrt{H}$-behavior is observed in the whole $H$-regime below $H_{c2}$, which is determined by the resistivity.  As shown in Figs.\,2(d) and 3(d), large $C/T$ at $H=0$ indicates that the substantial number of quasiparticles are excited even at $T/T_c\approx 0.1$. 
Since entropy balance imposes $\int_0^{T_c}\{(C_e/T)-C_n/T_c)\}dT=0$, $C_e/T$ for $x = 0.20$ is expected to decrease rapidly with decreasing $T$ below 0.4\,K.  Therefore largely remained $C/T$ arises from the Fermi pockets with extremely small superconducting gap.  The $H$-dependence of $C(H)/T$ for $x=0.20$  suggests the presence of Fermi pocket(s) with very small gap and other pocket(s), whose gap is larger and highly anisotropic.   Thus the gap structure in the tetragonal regime appears to be essentially different from that in the nematic regime.

  
  \begin{figure}[b]
 	\begin{center}
 		\includegraphics[width=0.98\linewidth]{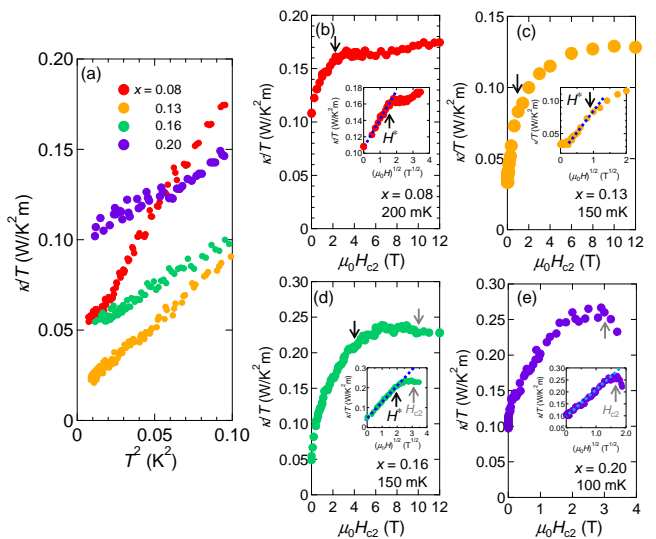}
 	\end{center}
 	\caption{(a) Temperature dependence of thermal conductivity divided by $T$, $\kappa/T$, plotted as a function of $T^2$ in zero field.  Magnetic field dependence of $\kappa(H)/T$ for (b) $x = $ 0.08, (c) 0.13, (d) 0.16 and (e) 0.20.  The gray arrows indicate $H_{c2}$.  The insets show $\kappa/T$ plotted as a function of $\sqrt{\mu_0 H}$. $H^*$ represents a magnetic field at which $\kappa(H)/T$ deviates from the $\sqrt{H}$-dependence.  The black arrows in the main panel indicate $H^*$.  }
 	\label{fig:Fig3.eps}
 \end{figure}

The thermal conductivity provides additional pivotal information on the superconducting gap structure, because the heat transport detects only the delocalized quasiparticles, insensitive to the localized quasiparticles. Figure\,4(a) depicts $\kappa/T$ plotted as a function of $T^2$ in zero field. 
At low temperature, $\kappa/T$ is well fitted by $\kappa/T=\kappa_{0}/T+bT^2$, where $b$ is a constant. 
We confirmed that the ratio of $\kappa_{0}/T$ and the electrical conductivity $\sigma_0$ at $T \rightarrow 0$ above $\mu_0H_{c2}$ is $(\kappa_{0}/T)/\sigma_0=(1.04 \pm 0.02)L_0$ for $x=0.16$ and 0.20, where $L_0=\frac{\pi^2}{3}\left(\frac{k_B}{e}\right)$ is the Lorenz number, indicating that the Wiedemann-Franz law holds. 
At zero field, the presence of a residual value in $\kappa/T$ at $T\rightarrow 0$, $\kappa_{00}/T$, indicates the presence of normal fluid, which can be attributed to the presence of line nodes in the gap function.  Finite $\kappa_{00}/T$ is clearly resolved in $x=0.08$, 0.16, and 0.20, indicating the presence of line node.  On the other hand, $\kappa_{00}/T$ for $x=0.13$ is much smaller or vanishes at $T\rightarrow 0$.

Figures\,4(b), (c), (d) and (e) depict the $H$-dependences of $\kappa(H)/T$ for $x=0.08$, 0.13, 0.16 and 0.20, respectively.  Similar to $C(H)/T$,  the application of small magnetic fields  causes a steep increase of  $\kappa(H)/T$ for all $x$; as shown in their insets, $\kappa(H)/T$  increases with  $\sqrt{H}$ at low fields.   Same as the specific heat, the $\sqrt{H}$-dependence of $\kappa(H)/T$ appears as a result of Doppler shift of quasiparticle spectra in the presence of line nodes.  For $x=0.08$, 0.16 and 0.20, $\kappa(H)/T$ increases immediately when the magnetic field is applied.  (We note that the lower critical field $H_{c1}$ is much smaller than the field scale of interest.~\cite{Abdel-Hafiez15})  This $H$-dependence, along with the presence of finite $\kappa_{00}/T$, indicates the presence of line nodes.   For $x=0.13$, on the other hand, $\kappa/T(H)$ is insensitive to $H$ at very low fields even above $H_{c1}$, suggesting that although the gap function has deep minimum at certain directions, it is finite, i.e. no nodes.  This is consistent with very small or absence of $\kappa_{00}/T$.  As shown in the insets of Figs.\,4(b), (c) and (d),  $\kappa(H)/T$ deviates from the $\sqrt{H}$-dependence above $H^*$ for $x=0.08$, 0.13 and 0.16.  The values of $H^*$ for $x=0.08$ and 0.13 are close to the ones observed in $C(H)/T$ in Fig.\,3(b) and (c). Above $H^*$, $\kappa(H)/T$ shows much weaker $H$-dependence than below $H^*$.  In particular, $\kappa(H)/T$  is nearly $H$-independent  for $x=0.08$ and 0.16.   Since thermal conductivity is insensitive to localized quasiparticles inside vortices, $\kappa(H)/T$ in fully gapped superconductors is independent of $H$ except for the vicinity of $H_{c2}$. Thus the observed initial steep increase with $\sqrt{H}$-dependence, followed by much weaker $H$-dependence of $\kappa(H)/T$, provides evidence for the multigap superconductivity, in which  small gap has line nodes or deep minima and large gap is nearly isotropic~\cite{Watashige17}.  This is consistent with the conclusion drawn from the specific heat.  For $x=0.20$, $\kappa/T$ increases with $\sqrt{H}$ nearly up to $H_{c2}$, which is again consistent with the specific heat.

Next, we compare our results with other experimental observations. 
The observed $\sqrt{H}$-dependence of $C(H)/T$ and $\kappa(H)/T$ only at low fields for $x< 0.08$ is consistent with the results of ARPES and QPI for $x<0.07$~\cite{Xu16,Hashimoto17,Sprau16}, showing the large gap anisotropy in the hole pockets. This initial $\sqrt{H}$-dependence persists in the whole nematic regime. 
Therefore it is natural to consider that the superconducting gap in the hole pocket is always highly anisotropic in the whole nematic regime.   
The gap structure of the electron pockets has been less clear.  
In fact, no gap has been observed on the electron pockets in ARPES measurements~\cite{Xu16}. 
However, the fact that $H^*$ is much smaller than $H_{c2}$ implies that the gap of the electron pockets is  larger than hole pocket. 
Moreover, $H$-dependences of $C/T$ and $\kappa/T$ above $H^*$ suggest that the gap of the electron pockets is much more isotropic than the hole pocket. It should be stressed that the line nodes in the hole pocket are accidental, not symmetry protected, because as directly revealed by the STM measurements, the nodes are lifted near the twin boundaries~\cite{Watashige15}.  Moreover, the presence of line nodes has been reported by thermal conductivity measurements on some crystals for $x=0$ \cite{Kasahara14}, while small but finite gap has been observed in different crystals~\cite{Hope16}, which may be attributed to the difference in the amount of impurities and twin boundaries.

 Since the elliptical hole pocket becomes more circular with increasing $x$, the highly anisotropic gap in the hole pocket in the whole nematic regime implies that the anisotropic pairing interaction is little influenced by the elliptical distortion of the hole pocket. This immediately excludes the possibility of the intraband pairing, in which the superconductivity is mediated by fluctuations with very small momentum.  This is because in such a case the gap anisotropy should be sensitive to the shape of the Fermi surface.  As displayed by green area in the hole pocket in Fig.\,1(a), the gap node/minimum locates at the area dominated by $d_{xz}$ orbital character for $x=0$ and 0.07~\cite{Xu16,Hashimoto17,Sprau16}. To explain such a highly anisotropic gap in a tiny Fermi pocket, a pairing interaction which is strongly orbital dependent has been proposed~\cite{Kreisel17, Yamakawa16}. In this scenario, the gap minimum/node appears as a result of the strong nesting properties of $d_{yz}$ orbit area, shown by red in Fig.\,1(a), between the hole and electron pockets. Since the pairing interaction is dominated by $d_{yz}$ orbital, the gap minimum/node can appear in the area with $d_{xz}$ orbital character in the hole pocket. 
In fact, strong nesting properties between $d_{yz}$ orbitals has been discussed in BaFe$_2$As$_2$ with stripe type magnetic order.     

Finally we discuss the superconducting gap structure in the tetragonal regime.  In the orbital dependent pairing scenario, the node/minimum is expected to appear 
the portion of the electron pockets (e1, e2) with $d_{xy}$ character [shown by blue in Fig.\,1(b)], 
because $d_{xz}$ and $d_{yz}$ orbits in the hole pockets (h1, h2) are identical in tetragonal lattice.  This scenario suggests that the superconducting gap of hole pocket is very small, giving rise to substantially large $C/T$ even at $T/T_c\approx 0.1$.  

In summary, the thermal conductivity and specific heat measurements on FeSe$_{1-x}$S$_x$ in a wide $x$ range reveal  the presence  of deep minima or line nodes in the superconducting gap function both in the whole nematic and tetragonal regimes.  Moreover, the multigap nature of the superconductivity is commonly observed in both regimes.  These results imply that the pairing interaction is significantly anisotropic both in the nematic and tetragonal regimes.  However, we point out that the gap structure, particularly the position of the gap minima/nodes, in the tetragonal regime is essentially different from that in the nematic regime. Thus determining the positions of gap minima/nodes in the whole nematic and tetragonal regimes should provide a clue to understanding a pairing mechanism of highly unusual superconductivity in FeSe.

We thank T. Watashige for experimental support, and T. Hanaguri and H. Kontani for helpful discussion. This work was supported by Grants-in-Aid for Scientific Research (KAKENHI) (No. 25220710, No. 15H02106, No. 15H03688), Grants-in-Aid for Scientific Research on Innovative Areas ``Topological Materials Science" (No. 15H05852) from Japan Society for the Promotion of Science (JSPS).

\end{document}